# Maunakea Spectroscopic Explorer (MSE) Instrumentation Suite

Alexis Hill[1][a], Nicolas Flagey[a], Alan McConnachie[a], Rick Murowinski[a], Kei Szeto[a],

[a] CFHT Corporation, 65-1238 Mamalahoa Hwy, Kamuela, Hawaii 96743, USA

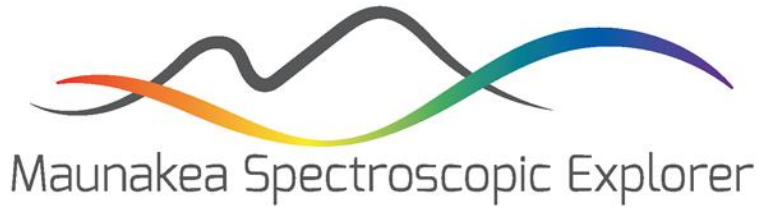

## ABSTRACT

The Maunakea Spectroscopic Explorer (MSE) is replacement of the existing 3.6-m Canada France Hawaii Telescope into a dedicated wide field highly multiplexed fiber fed spectroscopic facility. MSE is capable of observing over four thousand science targets simultaneously in two resolution modes. The paper describes the unique instrument system capabilities and its components starting from the telescope prime focus and ending at the spectrograph suite. The instrument system components have completed their conceptual designs and they include a Sphinx positioner system, fiber transmission system, low/moderate resolution and high resolution spectrographs and a calibration system. These components will be procured separately and the Project is responsible for their integration and the overall system performance afterward. The paper describes from a system perspective the specific design and interface constraints imposed on the components given the extra interface and integration considerations.

## 1. INTRODUCTION

Maunakea Spectroscopic Explorer (MSE) is an upgrade of the 3.6-meter Canada France Hawaii Telescope (CFHT) into a dedicated optical and near-infrared (NIR) spectroscopic survey facility.

MSE is a powerful survey machine with a 10-meter effective aperture telescope and a 1.5 square degree field of view located at its prime focus. MSE will operate in the optical to near-infrared, at low resolution (LR) (R = 2500), moderate resolution (MR) (R = 6000) and high resolution (HR) (R = 40,000). The instrumentation suite includes all hardware needed to collect the light at the prime focus, transmit it through the observatory to spectrograph detectors and calibrate it so that raw data from millions of targets per year can be collected and distributed.

MSE's first-light instrumentation takes advantage of Maunakea's excellent site seeing, large aperture and field of view by implementing a highly multiplexed fiber-fed system capable of collecting tens of thousands of spectra each night. The multiplexing includes an array of 4,332 fiber optics that are simultaneously precisely positioned to maximize the light entering them. The fibers are routed through the observatory to transmit the light to two banks of spectrographs several tens of meters away.

MSE's instruments are not unique in themselves, and in fact follow MSE's strategy of using technologies that are proven in operation as much as possible. However the instruments of MSE, with the overall architecture, combine to create capabilities that are not found currently anywhere in the world. This results in an excellent survey speed due to the combination of multiplexing and sensitivity, dedicated operations and flexible queue-based scheduling.

MSE has recently completed its Conceptual Design Phase (CoDP) [1], in which the architecture and design of the observatory and its instrumentation subsystems has been conceived, top level requirements have been defined and areas of future work for the Preliminary Design Phase (PDP) have been identified. During CoDP, MSE's instrumentation suite was developed during CoDP by partners in Australia, Canada, France and China. This paper highlights their current state of development, as well as the expected challenges and risks that will be resolved going forward.

---

[1] Email: hill@cfht.hawaii.edu; Telephone: 808-885-3187

## 2. MSE OVERALL CONFIGURATION

MSE is an altitude-azimuth prime focus telescope with corresponding elevation and azimuth structures to support the telescope optics and hardware to enable observations. The overall layout of the observatory is shown in Figure 1.

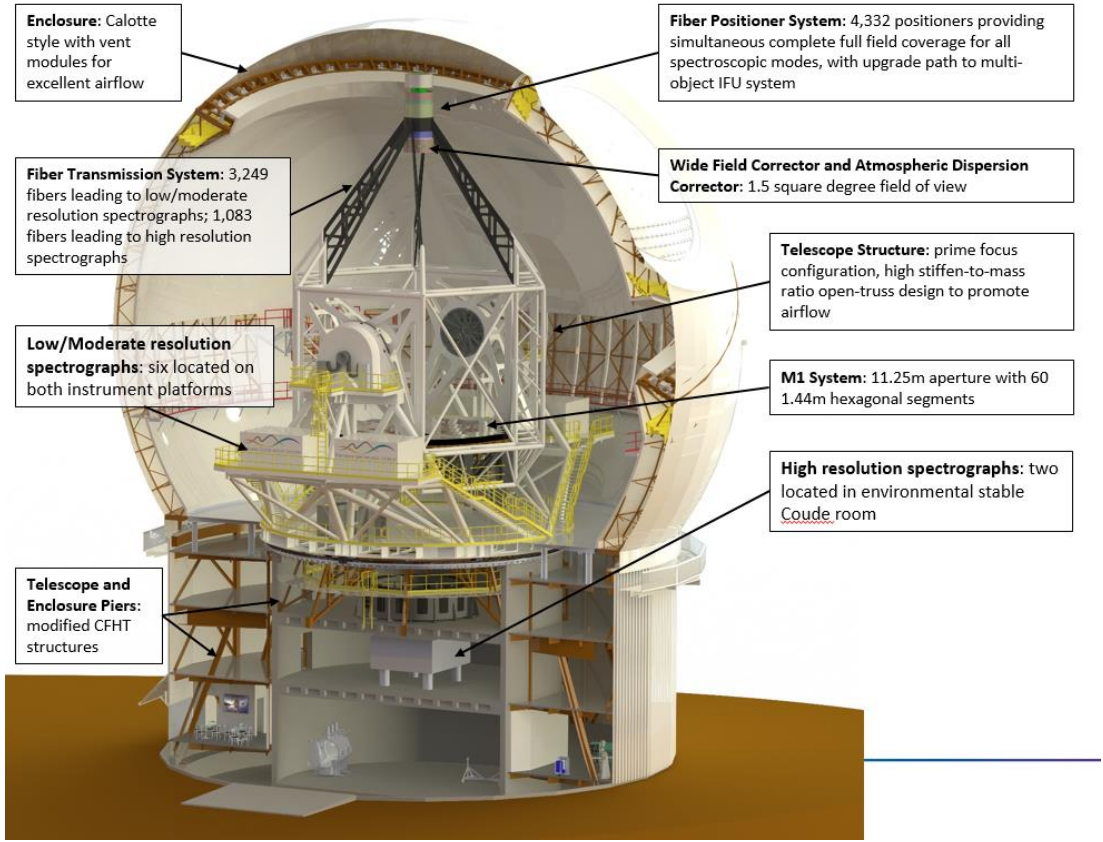

Figure 1. MSE overall layout after Conceptual Design Phase.

The elevation structure supports M1 (in a cell support structure) and WFC/ADC (at the top end of the telescope) in a stable configuration as the telescope moves through 0° to 60° Zenith motion. The azimuth structure rotates over ±270° and supports the elevation structure as well as instrument platforms on both sides of the structure.

The telescope includes a 60-segment primary mirror (M1) with an 11.25-m entrance pupil (10-m effective diameter) and a five-element wide field corrector/atmospheric dispersion corrector (WFC/ADC). M1 has 18.81 m focal length and radius of curvature is 37.698 m. This optical configuration delivers f/1.9 at a convex focal surface (Figure 2) with a radius of curvature of 11.33 m and a 1.52 degrees$^2$ field of view (584 mm diameter).

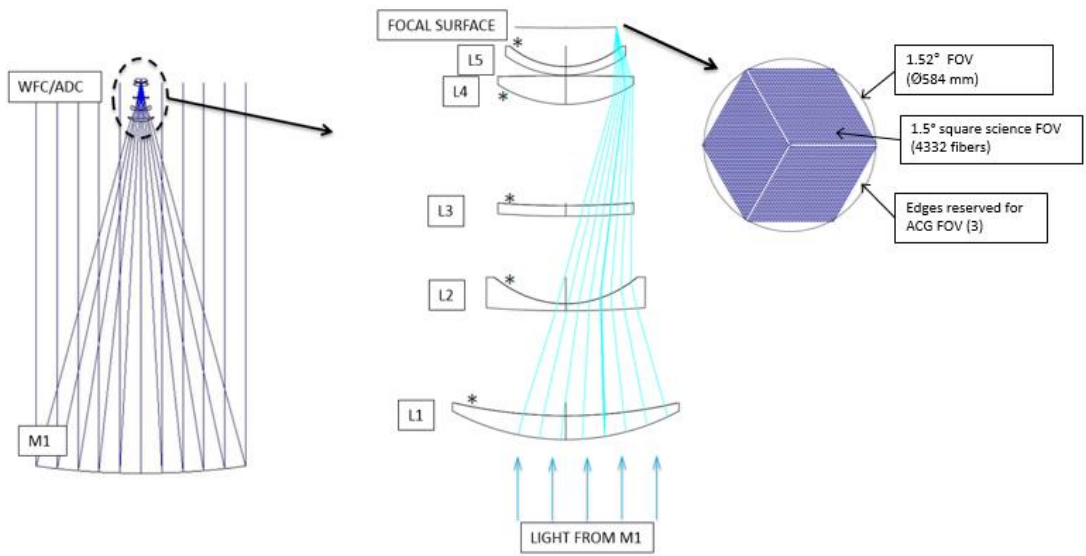

Figure 2. M1 and WFC/ADC (left, middle) and field of view at prime focus (right).

PFHS provides top end subsystem positional correction in five degrees of freedom (focus, decenter and tip/tilt) to compensate for dimensional changes of the telescope structure due to environmental and gravity orientation effects. By making these moves, PFHS maintains the alignment of the WFC barrel to M1, ensures the fibers and the guide cameras are positioned at the focal surface and provides a small offset as part of the ADC control action to allow for atmospheric dispersion correction. WFC/ADC is a single barrel opto-mechanical assembly, mounted on a Prime Focus Hexapod (PFHS).

For survey efficiency, the science field must be suitable for tiling so MSE has chosen to define the science field of view as a hexagon, taking up 1.5 degrees$^2$ (Figure 2). The hexagon is packed with fibers to collect light from targets and transmit them to the spectrographs. The remaining edges of the field of view are reserved for three off-axis guide cameras. Science targets in the field of view rotate as the telescope follows the sky, so the instrumentation and the guide cameras ride on a large-bearing instrument rotator (InRo) mechanism. Details of many of the subsystems at the top end of the telescope are available in [2] and are therefore not included here.

MSE's instrumentation suite (Figure 3) includes several subsystems, each designed by a partner institution.

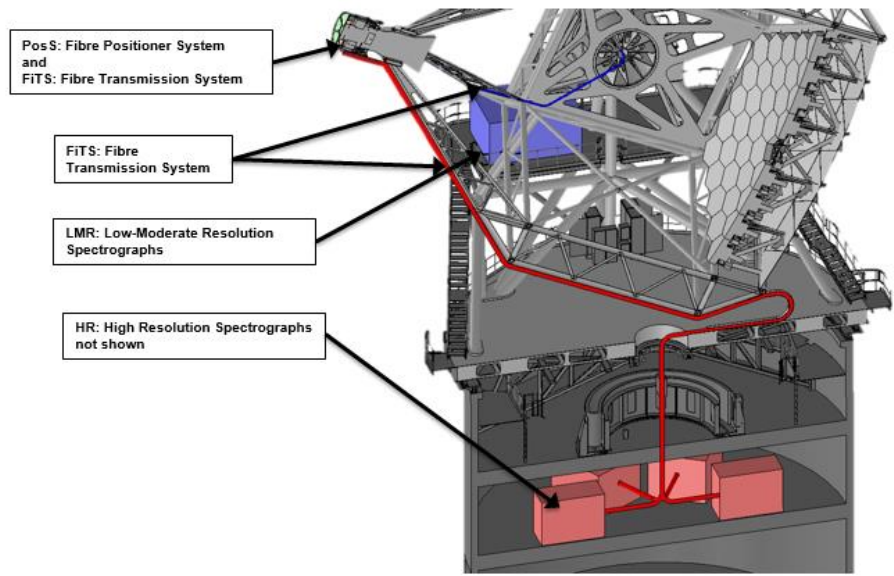

Figure 3: MSE instrumentation.

The Fiber Transmission System (FiTS) (provided by Herzberg Astronomy and Astrophysics, HAA in Canada, [3] includes more than 4,332 fibers that collect the light at the focal surface and deliver it to the spectrographs, several meters away. Along with providing the fiber bundles, FiTS subsystem also includes a fiber management system to accommodate field rotation during observations and routes and protects the fibers through all motions of the telescope in all environmental conditions.

The Fiber Positioner System (PosS) (provided by Australian Astronomical Observatory, AAO, see [4]) is an array of identical actuators, which carries and positions each fiber to a unique lateral position on the focal surface. To configure for each observing field, the positioners operate in closed loop with a dedicated metrology system (mounted in the missing central segment of M1) to maximize the light from targets entering the fiber.

The fibers transmit light to two banks of spectrographs, High Resolution (HR) and Low-Moderate Resolution (LMR), which collect the raw data of tens to hundreds of thousands of raw spectra every night.

HR spectrographs (designed by Nanjing Institute of Astronomical Optics & Technology, NAIOT in China, [5]) are fed by more than 1,083 fibers on the 2nd floor of the telescope pier, with a fiber length < 50 m. Each spectrograph unit has spectral resolution of R = 40,000 (< 600nm) and R = 20,000 (for > 600 nm) and operates over the range of 360-900 nm, with spectral channels (or wavelength windows) defined as: Blue [360-440] nm, Green [420-520] nm and Red [500-900] nm. The working windows are chosen to be Blue $\lambda/30$ at 408.55 nm, Green $\lambda/30$ at 481 nm and Red $\lambda/30$ at 650.5 nm. The working windows can be changed at any time by changing the dispersers.

LMR spectrographs (designed by Centre de Recherche Astrophysique de Lyon, CRAL in France, [6]) are fed by 3,249 fibers on the instrument platforms on the telescope's azimuth structure, with a fiber length of < 35 m. LMR are a bank of six identical spectrographs each with three optical (blue, red and green) and NIR/H-band and switchable between low (LR = 2500) and moderate (MR = 6000) resolutions.

A system of calibration sources (SCal) is planned which includes a dome flat screen and light source on the enclosure and telescope structure, respectively, and light sources under the spiders for lamp flats and arcs. While a calibration plan has been developed [7], the specific requirements and implementation of the SCal system are a work in progress.

A multi-object integral field unit (IFU) is planned [8] as a second generation instrument upgrade, after some years of survey operations. However, the design is modular in that PosS can be replaced by the IFU with minimal alteration.

MSE is designed with reference to performance, operational and other constraints. Performance requirements are those that affect the performance of MSE, defined as sensitivity in the SRD [9]. Sensitivity is quantified by its signal to noise ratio (SNR), made up of quantities Throughput, Noise, Injection Efficiency and Image Quality, all of which are discussed in detail in other literature [10], [11]. As well, MSE's operational concept [12] introduces additional constraints on the instruments, especially those related to calibration. These are discussed as needed in the following sections of this paper.

## 3. INSTRUMENTATION

### 3.1 Positioner System (PosS)

For a given image quality delivered to the focal surface, injection efficiency is dependent on the accuracy of the positioners (both laterally and in the focus direction) and tilt angles of the positioners (which affects defocus). After parallel development of two positioner technologies (phi-theta and tilting-spine), MSE has selected "Sphinx" tilting-spine robotic actuators (Figure 4) as the baseline design. It was determined through modeling of Injection Efficiency [13] that the technologies have very similar performance, despite added defocus errors caused by the tilting spines given other sources of positional errors are included.

Selection was based on other factors. One of the critical factors was the flexibility and multiplexing that is enabled by having simultaneously HR and LMR full field coverage which significantly improves the overall observing efficiency [14] of MSE. As well, there are concerns that phi-theta positioners require optical switches because they must carry one each of a HR and LMR fiber. This would have been a source of throughput losses and mechanical failure. Other benefits of the Sphinx tilting spine system include low dissipated heat near the focal surface and proven long-term reliability compared to phi-theta technology.

Importantly, the choice of tilting spine actuators is thought to induce minimal stress as fibers are moved to position, which minimize transmission losses and throughput variations due to bends in the fibers. This also creates changes in FRD and near-field and far-field effects, affecting wavelength resolution. This will be measured on a fiber test bench being developed for MSE by the University of Victoria, in Canada [15].

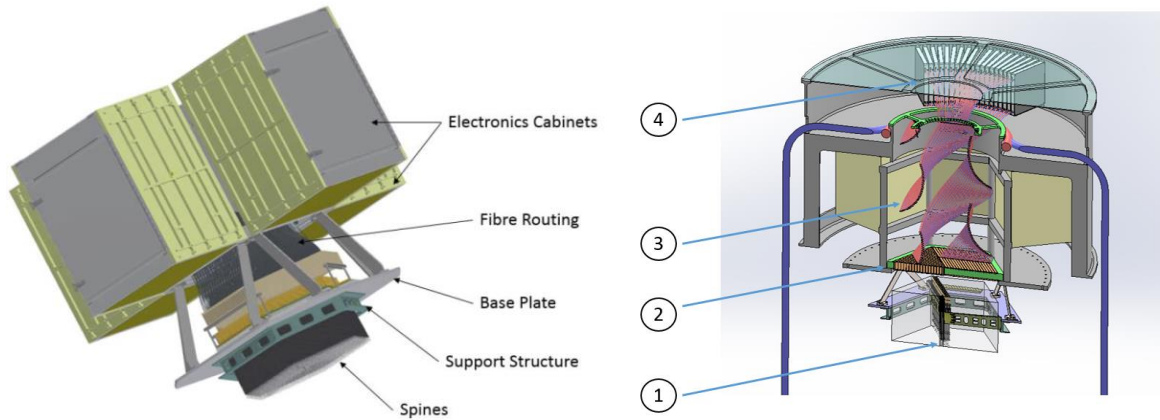

Figure 4. PosS and FiTS and the top end. PosS (left) and FiTS (right): 1) Positioning system (simplified), 2) fiber combiner, 3) helical tubes, 4) loop boxes

The Sphinx system includes 4,332 individual actuators, each closely integrated with a fiber from the FiTS system. The piezo actuator "spine" assemblies (Figure 5) include several components. Carbon fiber tubes house the FiTS fiber (with ferrule). The tubes are supported by a pivot ball, held in place by a magnetic "cup" which acts as the fulcrum of the spine. A piezo actuator produces a stick-slip of the cup and ball interface, which translates into an angular displacement of the spine. The resulting small lateral movements of the fiber end the focal surface are accurate to within 6 um RMS when used in closed loop with the metrology system during configuration. When the spine is tilted about the pivot ball to its full patrol range, the defocus of the fiber tip versus the fiber tip at vertical is 80 um (max). Since the amount of defocus will vary from positioner to positioner, PFHS will position the system to correspond to the median of all of the spine tilts. However, in modeling of the IE, the tilt distribution per field has been modeled [18] for which the defocus is smaller than the median tilt so the IE is actually conservative in this regard but nonetheless meets science requirements.

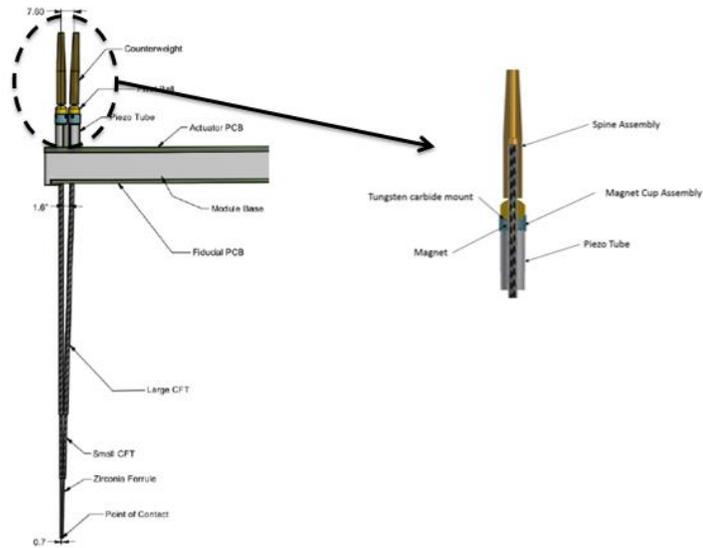

Figure 5. Fiber positioner spine assembly.

To aid assembly and maintenance, spine actuators are mounted in manageable modules (Figure 6), which are integrated on a stable reference structure (not shown), which is integrated/aligned as closely as possible to the theoretical curved focal surface created by the optical system.

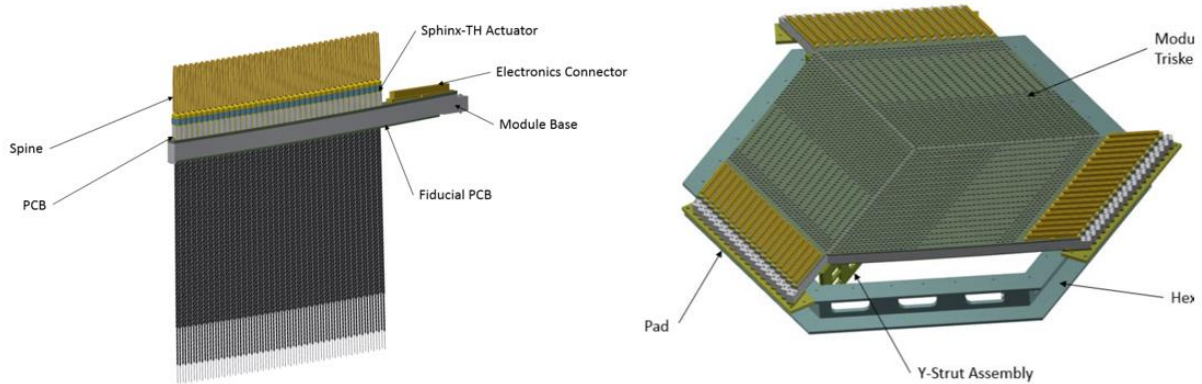

Figure 6. Fiber positioner modules.

The pitch of the actuators is 7.7 mm, with each actuator capable of moving the tip of each spine within a field patrol radius of 1.24 times the pitch (9.63 mm). The patrol areas overlap, allowing 1,083 HR and 3,249 LMR fibers to maintain full field coverage, with 97% of field positions accessible by 3 or more LMR fibers and 58% of field positions accessible by 2 or more HR fibers. Fiber to fiber collisions are low risk as they will not cause damage to the fibers or actuators. A metrology system images the full array of positioner fibers and iteratively works in closed loop with the positioner system to achieve its accuracy during configuration.

The Sphinx design is an evolution of the piezo-actuated technology, first designed and implemented in FMOS-Echidna (Subaru), and later refined and simplified through design studies and prototypes for various other systems. Sphinx represents a mature and low risk solution to the MSE's positioner requirements.

## 3.2 Fiber Transmission System (FiTS)

The design of FiTS is primarily driven by throughput (FRD and near- and far-field) requirements and the need to have stable and repeatable calibration over 24 hours (according to the current calibration procedure in the operations concept [12]), over the full range of motion. This includes throughput due to transmission losses (mostly based on fiber length and use of continuous fibers), Fresnel (input and output) losses and focal ratio degradation (FRD). As well, the entire subsystem must be robust to prevent fiber breakage over the range of motion of the telescope and allow for repair in the event fibers are damaged over their lifetime.

To provide the highest possible throughput, fibers are provided in a continuous link, using no connectors, all of the way from the focal surface to the spectrograph inputs. After exiting the PosS actuators, the FiTS system support the fibers with protective tubes before being wound in a helix in a Rotation Guide System to allow for the motion of InRo. The fibers are encased in bundles and fed through the telescope through all of its altitude and azimuth rotations. The fibers are routed to both the instrument platforms and the 2nd floor spectrograph room (currently a Coudé room in CFHT, see [15]).

High (0.26 – 0.28) numerical aperture fibers capable of accepting f/1.9 have been selected to avoid adding additional input optics at the fiber input and the resulting throughput losses. During CoDP, it was found that FRD in this type of fiber is relatively small. The fibers are expected to include an anti-reflection coating on the input and output ends, however the feasibility of applying the coatings is a subject of future work.

Fibers are susceptible to FRD, which has been estimated and tested in CoDP for this particular type of high NA fibers. Future work includes testing based on the stability of FRD which will affect the calibration of the system. As mentioned previously, a test bench has been built for this purpose [16].

The HR fiber diameter is 0.80" and the LMR fiber diameter is 1.0" and are sized [17] to take advantage of delivered image quality at the focal surface, based on the natural seeing distribution on Maunakea, dome seeing and the optical design. The fiber size for HR is undergoing a trade to optimize the SNR overall versus practical constraints on the spectrograph.

Throughput is also affected by the length of the fibers, particularly at the blue end of the wavelength spectrum. The current baseline of having the HR spectrographs in the inner pier (< 50 m fiber length) and the LMR spectrographs on instrument platforms (< 35 m fiber length) is the subject of a trade study for MSE as it has been determined that the HR spectrograph has the potential to benefit much more from the shorter fiber length in the blue wavelengths than the LMR spectrograph. This has the potential to improve the overall efficiency of MSE.

At the spectrographs, fibers are terminated in slit input units that provide the interface from the fibers to the spectrograph. The interface to the spectrographs is required to have a spherical shape or "smile" due to the off-axis collimators in both spectrograph designs. The shape of the slit compensates for the optical distortion such that spectra are "flat" (or straight line) when delivered to the spectrograph detectors. It is expected that these will look something like the slit input unit from Hermes (Figure 7) but with the design features, such as V-grooves and strain-relief proposed by HAA (Figure 7).

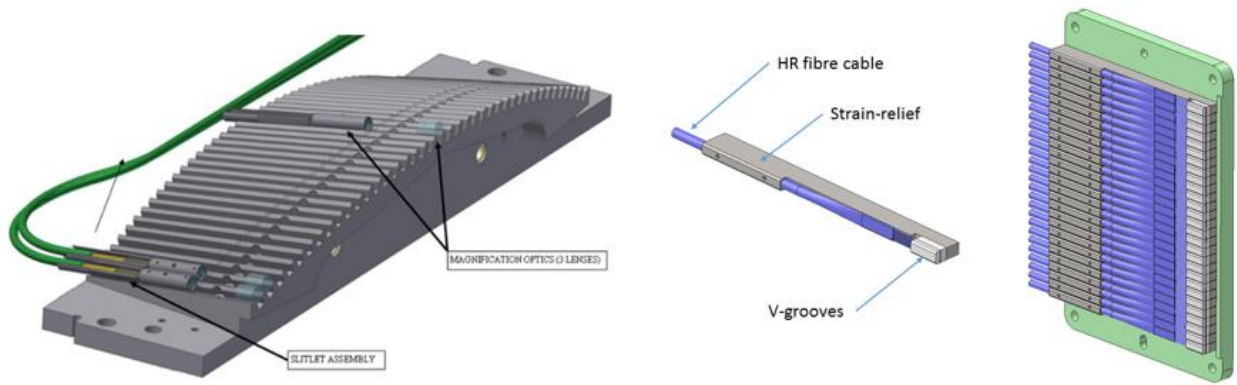

Figure 7. Left: Curved slit input unit from AAT-HERMES (shows magnification optics that will not be included in MSE). Right: Straight slit input unit proposed by HAA.

Apart from operation at very high NA fiber with fast f/ratios, most of the aspects of the FiTS design will be very similar to other multi-object fiber-fed projects currently underway. One major challenge is industrializing the fabrication process to build cables for thousands of fibers and finding a reliable process to AR-coat the ends of the fibers. This will be addressed in PDP.

The integration of a long continuous fiber length, creates integration, maintenance and presents repair risks. Fibers cannot break and must be robust and field repairable. Current thinking is to employ fusion bonding for this purpose. In CoDP, FiTS determined it is possible to achieve good performance and consistency using fusion bonds in a controlled condition. Whether this can be adapted to an in situ repair process is a subject for future work.

### 3.3 High Resolution Spectrograph System (HR)

Based on science goals, the requirements that drive the design of the HR spectrographs are throughput, noise and spectral resolution. The optical design by NAIOT takes these quantities into account.

The overall optical off-axis layout is shown in Figure 8. Two f/2 collimators were compared, an on-axis Schmidt or off-axis Houghton, with the off-axis design being adopted as the baseline as the throughput loss from the on-axis design was unacceptable.

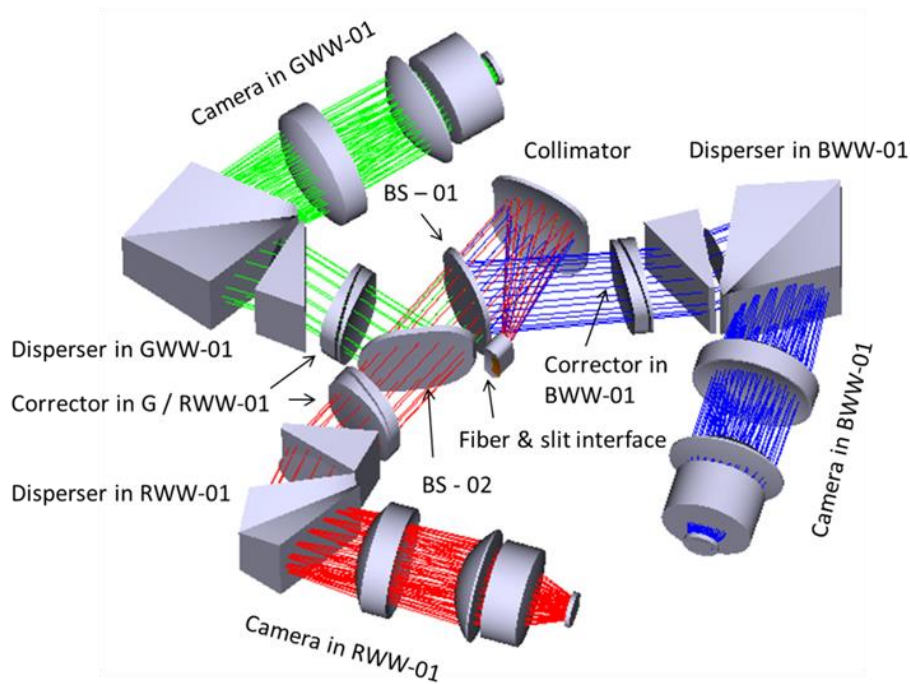

Figure 8. HR optical layout.

HR are fed light through slit input units, each of which secures the 0.80" core fibers, arranged in a spherically curved slit (for the off-axis optical arrangement). Fiber diameter directly affects the resolution and MSE may decide to reduce to a 0.75" (in future work) if this is feasible considering a resulting loss in injection efficiency.

Each of the three optical arms has its own camera with a maximum clear aperture of 500 mm and total physical length of 900 mm, feeding a CCD detector. The detector system sits on individual stages that provide focus adjustment over ±1 mm to accommodate working window changes in the future and are cooled via helium closed-cycle coolers. The cameras and collimator for the off-axis design have large aspheres which could be avoided by using an on-axis design. These are expected to be challenging to fabricate. NAIOT is working with vendors to determine feasibility, risk and cost for the aspheres.

MSE HR is challenging due to the 10m class large telescope aperture and its high resolution of 40,000 in the wavelength range of 360-500nm. This is driving the disperser (currently planned as a VPH grism and prism combination) to a challenging ultra- high line density (5,800 lines/mm) on a 700 mm x 400 mm substrate. As well, this constraint may be relaxed after review by the science team based on the science mission. This is future work as well.

## 3.4 Low-Moderate Resolution Spectrograph System (LMR)

Low and moderate resolution functions are combined into a bank of six identical spectrographs. The optical layout for LMR is shown in Figure 9. Each spectrograph has three optical arms (360 to 950 nm) and one infrared arm each with its own off-axis Schmidt camera, with dichroics between the collimator mirror and the corrector lenses. The spectrographs are independently switchable so that the visible arms can be switched between LR (R = 3000) or MR (R=6000). For low resolution, there is a corrector and VPH grating. For moderate resolution, the corrector and VPH grating are exchanged for a sapphire grism with immersed VPH. The science goals of spectral resolution and spectral coverage drive the LMR design, along with meeting sensitivity requirements by achieving the throughput and noise requirements.

The beam sizes are approximately 175 mm with f/1.2 cameras for each arm, including glued doublets of varying glass choices, depending on the camera. Each spectrograph is switchable between two main modes: optical (LR) plus J-band (LR) or optical (MR) plus H-band (LR). All arms of all spectrographs are independently controlled, allowing a great deal of flexibility in selecting simultaneous LR and MR modes for different targets in the same observing field. The limitation to this flexibility is only in the sophistication of the scheduling software.

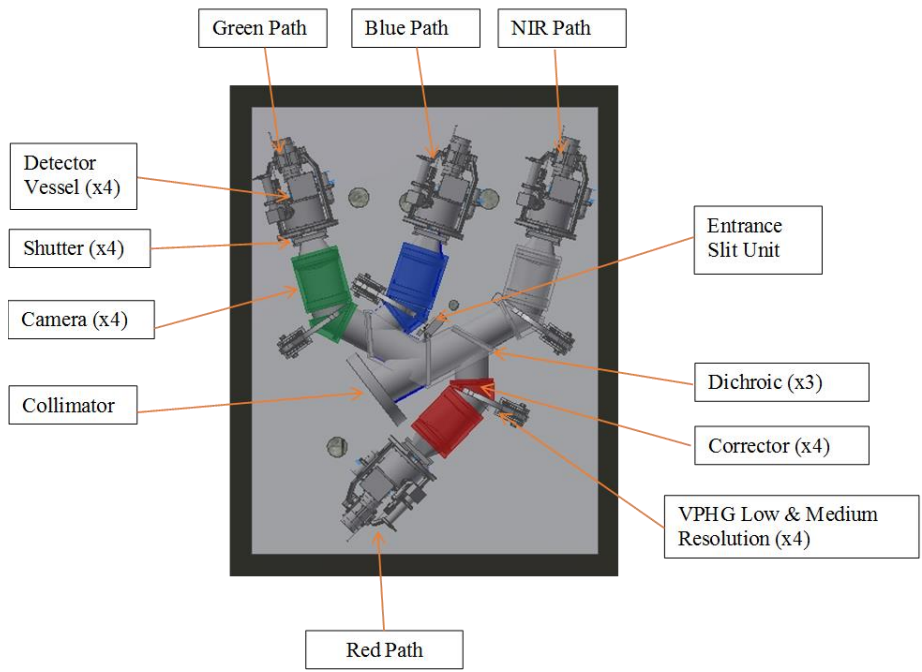

Figure 9. LMR optical layout.

The cameras are transmissive at f/1.2. Each arm has a single 61 mm x 61 mm detector, either and E2V 231-series CCD or a Teledyne Hawaii 4RG15. In both cases, the pixel size is 15 um detectors and are cooled with a linear pulse tube detector head.

The most significant challenges include fabricating the highly aspheric lenses, the tight space constraints in this concept of the optical design and the mechanical challenges of moving the dispersers via mechanisms. These challenges are based on manufacturability and maintainability rather than any particular performance requirements.

The design currently meets the throughput requirements however fabrication of the optics may be difficult and the throughput will be closely tracked through future phases. Spectral resolution is being met by the system but not in MR mode in the blue arm.

Noise based on optical background is closely linked to the instrument temperature and associated thermal emission of the whole spectrograph visible from the detector. Optical stability may be an issue, if thermal variations are too great or if mechanisms are not repeatable enough (particularly the grating change mechanism). Temperature changes will cause changes to the glass optical indices, grating line numbers and air space and overall lens positions and orientation. For calibration, the criteria is 0.045 of a resolution element which corresponds to 3.3 pixels of width at minimum. This corresponds to +/-1.1 μm which is estimated to require thermal control of +/-0.1°C. This could be mitigated by considering a simultaneous wavelength calibration on some of the fibers, however this is not currently in the plan.

H-band imposes a very cold instrument. Reducing background will require active cooling of the instrument to -70°C. This is at the limit of what can be achieved practically and a full vacuum cryogenics instrument is not desirable.

Future work will include consideration by the science team to reevaluate the H-band requirements and better define the calibration requirements. As well, the optical design will undergo some iteration to make it less risky and expensive overall, if possible and find an implementation of H-band that is not onerous.

### 3.5 Science Calibration System (SCal)

Calibration of the system is based on three main science-based requirements: velocity resolution accuracy, relative spectrophotometry and sky subtraction. This is discussed in more detail in a separate paper [7].

MSE plans to perform calibration exposures using both flats and arcs. SCal subsystem consists of those light sources needed to provide the flats and arcs.

Lamp flats will be taken in the nighttime before and after every observation under the same system configuration. Lamp flats will be taken in a fixed reference configuration to connect the twilight flats to the lamp flats. Twilight flats will be taken for a greater level of uniformity in reference configuration to characterize the overall system throughput. Daytime dome arcs will be taken in the fixed reference configuration. Daytime lamp arcs in the reference frame will be taken to connect the dome arcs to the lamp arcs to characterize the overall system wavelength. Nighttime lamp arcs will be taken before and after for the same system configuration.

The main driving constraints for SCal are in achieving sufficient sky subtraction at the faint limits MSE is designed to reach. This is highly dependent on several factors that are not related to the SCal system but it is clear that close attention to hardware design is needed in coordination with the calibration methodology.

Detailed SCal hardware requirements based on the calibration plan have not been defined yet and therefore this system has not undergone CoDP. This is a high priority in the near future for MSE.

### 3.6 Integral Field Unit (IFU)

An IFU [8] is planned as a second generation capability for MSE. It is anticipated that at a minimum, the existing PosS system can be replaced with an IFU that will use the existing LMR spectrographs. A second possibility is that the IFU could co-exist at the top end of the telescope with the positioner systems, be deployable and switching could be implemented at the LMR end. This initial concept assumes that overall science requirements on the first light systems, such as wavelength ranges and spectral resolutions, will be the same or similar as those for the IFU. However, the specific science cases for the IFU have not yet been defined and work must be done to determine the optimal configuration (for example, number of spectra per IFU versus number of IFUs). This will be developed by the science team as the project moves into PDP. In the meantime, the conceptual design of MSE has been developed to allow for this upgrade.

## 4. MSE SYSTEM INTEGRATION CONSTRAINTS

The subsystems of MSE are being designed and built by organizations distributed around the world, so interfaces between the subsystems, as well as detailed planning of integration at the observatory becomes a critical factor for the project to succeed in meeting its performance requirements, on time without delays and rework. The Project Office is responsible for integrating all of the systems once delivered to the telescope.

Assembly and integration considerations are a major focus for the FiTS system. FiTS/PosS can certainly be built as an assembly but the complication of having unbroken fiber length will require some careful planning about how the system is assembled and integrated at widely geographically dispersed facilities, shipped to the observatory and then installed on the telescope. The current plan is to build a continuous fiber bundle, deliver the entire subsystem to the PosS assembly facility as a unit, with bare fibers at the input end and then assemble them and add the slit input units for the spectrographs. The details of this interface are under development but certainly fibers will be mounted in a zirconium ferrule, the fiber end polished and AR coated and integrated with individual spines of PosS.

Fibers can be easily broken and installing tens of meters of delicate fibers on the telescope, with a positioner system on one end and a set of slit input units on the other end will require careful planning. The process of fusion bonding in situ to repair broken or damaged fibers is being developed and the performance of fibers after such a process is being verified in current testing facilities. The sequence and responsibility of this activity is not determined yet and is a high priority for MSE.

The AIV process is controlled so that alignment tolerances on mating components ensure that fibers are aligned with the focal surface. This includes all prime focus subsystems, such as InRo and WFC/ADC have closely controlled interfaces to avoid large tolerance stack-ups. The IE budget includes AIV tolerances and are achievable with careful planning and interface definition and control.

HR and LMR will be built and independently verified at their fabrication facilities before being shipped directly to the observatory site. Integration of the spectrographs is a well understood process, with the exception of handling and access in the tight spaces expected at the telescope. For FiTS and HR/LMR, the input slit unit tolerances and design will be closely controlled and monitored so they don't degrade performance in the system after integration.

As well, the instrumentation package is constrained by the space and mass limits both at the top end and, for the spectrographs, at their locations within the observatory. This will be closely tracked through the project lifecycles.

## 5. CONCLUSION

MSE has a combination of capabilities not available in any facility in the world today. The instrumentation suite is purpose-built to enable these capabilities: massive multiplexing, sensitivity as demanded by the science cases and dedicated operations.

MSE will observe over four thousand science targets simultaneously in two resolution modes using the 4,332 fibers with individual positioners. The spectrographs' size and numbers allow the collection of raw data from targets at a rate of millions of spectra per month.

Sensitivity is enabled by taking advantage of the excellent natural seeing on Maunakea and large aperture by making effective material and design choices in the fibers and spectrographs and by optimizing the injection efficiency of the positioner system. The Sphinx fiber positioners provide a robust and reliable solution for positioning accurately over the large field of view. Throughput is enabled by the system choices for material choices, minimizing surfaces and in minimizing the length of the fibers. Noise is managed and accounted for by careful design. Calibration will present a special challenge but is being carefully planned and will be developed in future work.

Dedicated operations over the wide range of resolutions and wavelengths combined with the flexibility of the spectrographs modes will mean the collection of. The spectrographs cover multiple spectral resolution with great flexibility because there are many different wavelength arms and resolution mode combinations available. Both HR and LMR are available simultaneously for any observing field due to the instrument suite design.

MSE's technical approach is to maximize use of existing designs and minimize development of new or unproven technologies. Even so, there are engineering challenges currently, including integration of the PosS and FiTS systems efficiently at the telescope, optical fabrication challenges in the spectrographs, especially in cost-effective fabrication of dispersers and aspheres. The result is an incredibly powerful, efficient and reliable survey machine that will be in high demand in the years to come.

## ACKNOWLEDGEMENTS

The Maunakea Spectroscopic Explorer (MSE) conceptual design phase was conducted by the MSE Project Office, which is hosted by the Canada-France-Hawaii Telescope (CFHT). MSE partner organizations in Canada, France, Hawaii, Australia, China, India, and Spain all contributed to the conceptual design. The authors and the MSE collaboration recognize the cultural importance of the summit of Maunakea to a broad cross section of the Native Hawaiian community.